\documentclass{elsart}
\usepackage{epsfig}
\usepackage{amssymb}

\newcommand{\be}{\begin{equation}}
\newcommand{\ee}{\end{equation}}
\newcommand{\beas}{\begin{eqnarray*}}
\newcommand{\eeas}{\end{eqnarray*}}
\newcommand{\bea}{\begin{eqnarray}}
\newcommand{\eea}{\end{eqnarray}}
\newcommand{\req}[1]{(\ref{#1})}
\begin{document}
\begin{frontmatter}
\title{Selection by pairwise comparisons with limited resources}
\author[frib]{Paolo Laureti},
\author[cop]{Joachim Mathiesen}, and
\author[frib]{Yi-Cheng Zhang},
\address[frib]{D\'epartement de Physique Th\'eorique, Universit\'e de Fribourg, Perolles CH-1700, Switzerland}
\address[cop]{The Niels Bohr Institute, Blegdamsvej 17, Copenhagen, Denmark}
\begin{abstract}
We analyze different methods of sorting and selecting
a set of objects by their intrinsic value, via 
pairwise comparisons whose outcome is uncertain. 
After discussing the limits of repeated Round Robins,
two new methods are presented: The {\it ran-fil} requires no previous knowledge
on the set under consideration, yet displaying good performances
even in the least favorable case. The {\it min-ent} method
sets a benchmark for optimal dynamic tournaments design.
\end{abstract}
\end{frontmatter}
%
\section*{Introduction}
In the Internet era the amount of available information is overwhelming:
the problem of finding and selecting the most relevant becomes, therefore,
crucial. Each selection operation is noisy, yet we want to estimate
the minimal amount of resources that is necessary to sort a large number
of items. Our study is a first step to establish a firm theoretical
bound for the new Information Theory introduced in \cite{csz03}, which
aims to give theoretical basis for evaluating and improving the current
and future search engines.

The method of paired comparisons has been extensively studied
by statisticians \cite{d88} and widely applied in various fields.
Usually one has to rank $N$ objects (or agents), each one of them endowed
with a scalar quality $q_i, i=1,2,...,N$, on the basis of a finite number
of binary comparisons $t_c$. The ``true'' rank $R_i$ of item $i$
is assumed to be zero if $q_i$ is the highest quality, one for
the second highest and so forth.
Upon assuming an \emph{a priori} probability distribution of quality
$\phi(q)$ and a given probability
distribution of outcomes of comparisons between two objects
of known quality $p_{i,j}$, it is possible to write the
joint likelihood of the outcomes as a function of individual qualities.
The best guess for the quality set is the one that maximizes the likelihood,
i.e. the one that would produce the given outcome with the highest
probability.

On the other hand scarce attention has been devoted so far to the problem 
of designing optimal tournaments with a limited number of games.
Round Robin (RR) tournaments, for instance, are common but not very effective,
because they assume that all comparisons are equally useful. In fact
one might be more interested in the upper part of the classment than
in its lower end and the result of some comparisons could be foreseen with
high precision on the basis of previous outcomes. This fact motivates
the filters we shall develop in the present paper. 
Let us first discuss the theoretical limits of repeated RRs with 
the following example. 

\section{Best selection via Round Robins}
We want to select the best out of $N=N(0)$ objects, by successive elimination, in $k_c$ rounds.
Round $k$ is completed once a RR among all the (surviving) $N(k)$ objects is performed; 
after each round $\Delta(k)$ objects are eliminated. Thus
$$
N(k+1)=N(k)-\Delta N(k).
$$
If the elimination were perfectly effective, after $k$
rounds the quality of the selected objects would be uniformly distributed
in the range $(1-\gamma(k),1)$, where $\gamma(k) \propto N(k)/N$.

Consider now two objects $i$ and $j$, with a difference in mutual
preference probabilities $\epsilon_{i,j}=|p_{ij}-p_{ji}|$, that are compared
$\nu$ times. Their average difference in points will be of order
$\nu\epsilon \pm \sqrt\nu$.
In many models of interest, like the Bradley Terry (BT) model
introduced below \req{pij}, the average $\epsilon$
of $\epsilon_{i,j}$ in the quality domain $(1-\gamma,1)$ is proportional to $\gamma$.
In this case, in order to have a significant separation of two agents,
they have to play at least
\begin{equation}
  \label{eq:21}
\nu \sim 1/\epsilon^2(k) \propto 1/\gamma^2(k) \propto N^2/N^2(k) 
\end{equation}
games on average. 
In addition to that, we should keep in mind that each object is compared
with the remaining $N-1$ in the first round.
If we want to keep a constant selecting power,
another factor $N/N(k)$ has then to be considered to account
for the diminishing number of opponents. In all, each object needs
be compared  $\nu \sim N^3/N^3(k)$ times before deciding if it survived
round $k$.

Now suppose that, at round zero, we eliminated $a$ percent of the 
original $N$ objects. At round $k$ we shall eliminate the same $a$ 
percent of the remaining $N(k)$ objects only in about $\nu$ rounds. Thus
$$
\Delta N(k) \propto a \left[ \frac{N(k)}{N} \right]^3
$$
If we take the continuum limit and solve the resulting differential
equation, this yields
\be\label{ennet}
N(t)=2 a N/\sqrt{t}.
\ee
The minimum number $n_c$ of comparisons needed for the champion to arise
can be calculated by integrating \req{ennet} from $0$ to
$k_c \propto N^2$. The result reads
\be\label{RRscal}
n_c\propto N^2 \log N.
\ee
In the remaining of this paper we will propose
two new procedures intended to improve this result.

\section{The ran-fil method}
We shall introduce an algorithm that can be used for finding an object with a large
quality, i.e. a highly ranked one, with no previous knowledge
of $\phi(q)$ and $p_{i,j}$. Inspired by a previous
work \cite{csz03}, it is constructed in terms of rounds. 
In each round we line up objects labeled 
with numbers $1,\ldots,N$ ($N$ assumed to be even) and
compare all the odd numbered ones with their successors of even
numbers: $1$ with $2$, $3$ with $4$ and so on.
A round ends once items $N-1$ and $N$ have been compared. 
Winners replace losers:
if the object at site $i$ is preferred over that at $i+1$, we replace the latter
with $i$, and vice versa. 
Before a new round starts all objects
are reshuffled. We define the time $t$ as the number of pairwise
comparisons made (excluding cases
where an agent is compared to himself).
\begin{table}
\begin{center}
\begin{tabular}{||c|c|c|c|c||}
          \hline
          \hline
        n& 12& 34& 56 & 78\\
        \hline
        start & $q_1q_2$& $q_3q_4$ & $q_5q_6$&$q_7q_8$\\
          \hline 
        1  & $q_2q_2$& $q_3q_3$ & $q_6q_6$&$q_8q_8$\\

          shuffle& $q_6q_3$& $q_3q_8$ & $q_8q_2$&$q_2q_6$\\
       \hline
       2 & $q_6q_6$& $q_3q_3$ & $q_2q_2$&$q_6q_6$\\
       \vdots &\vdots&\vdots&\vdots&\vdots\\
       \hline
       k& $q_6q_6$& $q_6q_6$ & $q_6q_6$&$q_6q_6$\\
      
       \hline
       \hline
 \end{tabular}
\end{center}
\caption{Example of a run of the ran-fil algorithm with no noise.}\label{table1}
\end{table}
Table \ref{table1} shows an
example of a run of our algorithm where, after $k$ rounds, $q_6$ is our 
designed winner, i.e. our guess for the agent with the best quality. 

Using the above approach, however,
there is a quite large probability of losing the item with the
highest quality in one of the initial rounds and therefore end up with a
poor average winner's rank $R_w$. In order to avoid this we introduce some noise
such that, in each round we reintroduce objects which have
been eliminated in earlier rounds. The noise level we use is fixed and
set by the number of agents $\eta$ we re-introduce in each round.
Agents gain one point for each time they have been preferred. Finally we
estimate the agent with the best rank, for a given time step $t_c$, as the
one who gained more points.

\subsection{Testing the run-filter on the BT model}
Let us now focus on the model
\be\label{pij}
p_{i,j}=\frac{q_i}{q_i+q_j},
\ee
first proposed by Zermelo \cite{z29} but often referred to as
Bradley-Terry (BT) model \cite{bt52}.
We tested the run-filter method on a population of agents
whose qualities were uniformly distributed between zero and one.
\begin{figure}
\centerline{\epsfxsize=3.0in \epsfbox{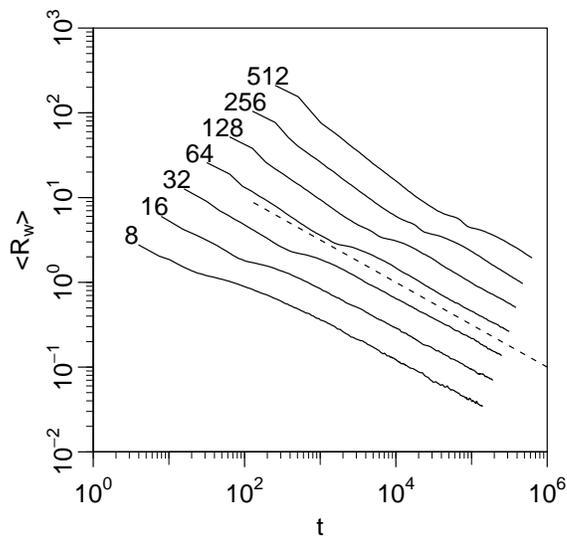}} 
\caption{The average rank $\langle R_w \rangle$ of the winning agent versus the number of
  comparisons in the run-filter method, for various values of $N$ (shown next to each line). 
The rank is averaged over $10000$
  runs and the noise level is put such that every second round an
  agent is re-introduced, i.e. $\eta=.5$. The dashed line is proportional to $1/\sqrt t$.}\label{fig17}
\end{figure}
Fig. \ref{fig17} shows the average rank of the winner as a 
function of the total number of comparisons $t_c$, for
different values of $N$ and with a reintroduction level $\eta=0.5$. For
the first few rounds our algorithm is similar to a knock-out
tournament, i.e. the number of teams in each round is halved. Note that
the ranking after some transient shows a characteristic scaling in the
number of comparisons $t_c\propto R_w^{\beta}$, and that the scaling
exponent $\beta$ seems to be independent of the number of agents. 

In the BT model the difference in mutual
preference probabilities of two agents with qualities $\alpha$ and
$\alpha-\delta$ ($1>\!\!>\delta>0$) is
\begin{equation}
  \label{eq:3}
\epsilon=  \frac {\alpha}{2\alpha+\delta}-\frac
  {\alpha-\delta}{2\alpha+\delta}=\frac \delta {2\alpha}+O(\delta^2).
\end{equation}
This means that when the surviving agents all have qualities
close to one, their average difference in $\epsilon$ scales almost
linearly with $\delta$ and equation \req{eq:21} holds. 
In the large $N$ limit
we can thus write down a differential equation for $\epsilon(t)$:
\begin{equation}
  \label{eq:15}
  \frac{d\epsilon}{dt}=-\alpha\epsilon^3,
\end{equation}
where $\alpha$ is a constant. The solution is explicitly written
as 
\be\label{scaling}
\epsilon(t)=\frac 1 {\sqrt{2\alpha}}t^{-1/2}. 
\ee
It now follows that
the estimated rank, $R_w(t)$, can be approximated by
\beas
   R_w(t) \approx \frac {N}{\epsilon(t)}\int_{1-\epsilon(t)}^1 (1-q) dq\nonumber=\frac 1 {\sqrt{8\alpha}}t^{-1/2}.\nonumber
\eeas
In fig. \ref{fig17} we have added a line with a scaling exponent of $\beta=-0.5$ in
the number of comparisons. The numerical examples are after some
transient in excellent agreement with the predicted scaling
\req{scaling}. 

The ran-fil algorithm is directly applicable when you want to estimate
the complete ranking table. In this case we define $\epsilon$ as  
$$
  \epsilon(t) =\frac 1 {N}\sum_{i=1}^{N}| \tilde R_i(t)-R_i|, 
$$
where $\tilde R_i(t)$ is the estimated rank of agent $i$ at time $t$
and $R_i$ its true rank. Following the same argument as above we
get the same scaling exponent. The prefactor, however, is larger. 

Note
that the noise is defined as simple as possible without any reference
to an underlying distribution of $p_{ij}'s$. One could, based
on \emph{a priori} knowledge of such a distribution, tune the noise to
increase the performance of our filter. Below we show that tuning of
the noise can improve performance dramatically.
\section{The min-ent method}
We shall outline here a method in which we assume 
that the fitness distribution $\phi(q)$
and the functional form of $p_{i,j}=f(q_i,q_j)$ are known a-priori. 
In real life it is very rarely the case, unless
these quantities can be reliably estimated from data
collected in the past. In some sports, for example, 
previous championships could provide such data.
Further inquiry
is needed to check how robust this method is with respect to 
errors in the existing knowledge. Similar problems
are widely dealt with in the literature 
and so we shall not tackle this question here.

The idea of our method is to
choose the couples to compare dynamically during the tournament.
At every time step a comparison $x(t)$ is performed and its 
outcome $w(t)$ recorded.
By time step $t$ each couple $(i,j)$ has played $n_{i,j}(t)\in(0,t)$ games. 
Let us denote by
$w_{i,j}(t)$ the number of times $i$ has beaten $j$ at time $t$, and by
$W(t) = ((w_{i,j}))$ the matrix of results collected until time $t$.
It follows that $w_{i,j}(t)+ w_{j,i}(t)=n_{i,j}(t) $ for each pair.

Let us now focus on the problem of finding only the best item.
Without any prior information it is natural to draw the
couples to compare in the first round
from a uniform distribution. Once matrix 
$W(t)$ is connected, though, we can adapt such a distribution so as to
maximize the acquisition of new information. The information provided 
by a new comparison $x_{u,v}$ can be quantified as
\be\label{info}
I_{W(t)}(x)=1-H\left(\psi|W(t) \cup x_{u,v}\right)/H_{max},
\ee
where $H(\psi | W(t)\cup x_{u,v})$ is the entropy of 
the conditional
distribution $\psi(k | W(t) \cup x_{u,v})$, the probability
that site $k$ has the highest fitness, given
the matrix of outcomes till time $t$ plus a new comparison $x_{u,v}$.
Ideally one would like such a probability 
distribution to be as picked as possible around the most probable 
value, which translates into information maximization.
This is a general procedure,
but different models require specific definitions of the
expected conditional information and of the weights characterizing
the importance sampling one wishes to apply \cite{Ma02}.

Here we shall test the procedure of comparing the maximizing
couple at each time step, i.e.
\be\label{max}
x(t+1)= Arg \max_{x_{u,v}} I_{W(t)}(x)
\ee
with the following definition of $\psi$:
$$
\psi(k | W(t) \cup x_{u,v})= p_{u,v} \psi(k | W(t) \cup w_{u,v}) + 
p_{v,u} \psi(k | W(t) \cup w_{v,u}).
$$
This corresponds to taking the expected value of
unknown outcomes.
Extension to the determination of the entire rank or
part of it can be easily found with the same reasoning.

\subsection{Finding the best item in a test model}
We test the outlined method on the model \cite{b54}
\bea \label{bechmod}
p_{b,j}&=&\pi >0, j \ne b \nonumber \\
p_{i,j}&=&0.5, i,j=1,2,...,N; i,j\ne b, \nonumber\\
\eea
which is the least favorable among many instances
\cite{ad96} and analytically solvable
in the case of Round Robins \cite{d88}.
Here, clearly, the assumption on a-priori distributions
translates into known win probabilities.
Thus
$$
\psi(k | W(t)) \propto \pi^{w_k} (1-\pi)^{n_k-w_k},
$$
where $n_k=\sum_j n_{k,j}$ and $w_k=\sum_j w_{k,j}$. 

We proceed as follows: at least $N$ comparisons are previously
made in such a way that the matrix of outcomes be connected. 
Then, at each time step $t$, we compute, for all possible couples $(u,v)$, 
the conditional entropy
\be \label{ch1}
H\left( \psi |W(t) \cup x_{u,v} \right) = 
\frac{
H( \psi |W(t)) + (1-a_{\pi}) ( \psi_u \log \psi_u +
\psi_v \log \psi_v ) - a_{\pi} \log a_{\pi} ( \psi_u
+ \psi_v ) }{ 
(1-( \psi_u + \psi_v ) (1-a_{\pi}) ) },
\ee
where  $a_{\pi}=2 \left[ 1- 2 \pi (1-\pi) \right]$ and $\psi_y$
stands for $\psi(y | W(t))$.
Next time step we shall
compare the couple $x(t+1)$ satisfying condition \req{max}.

We assign 
$\sum_{j \ne k} w_{k,j}/n_{k,j}$,
points to item $k$
and declare the winner as the one that collected more points
at time $t_c$. Then we check if our guess is right.
Notice that, although the above rule is arbitrary,
any other one would not improve notably our results \cite{d88}.
Even maximum likelihood estimations, which are the best ones 
under our hypothesis,
give the same ranking as ours in the case of Round Robins \cite{z29,f57}
yet involving much heavier calculations.

\begin{figure}
\centerline{\epsfxsize=3.0in \epsfbox{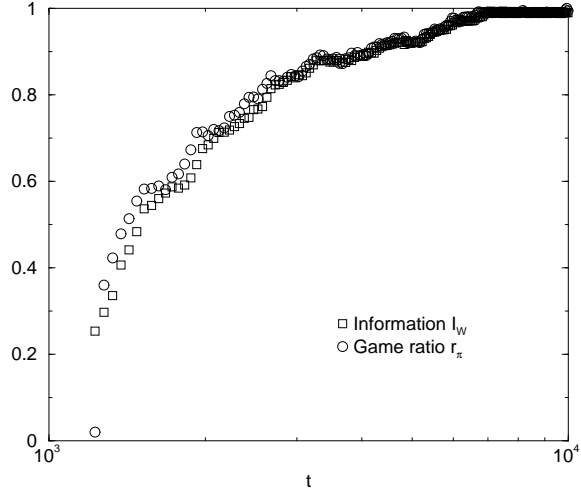}} 
\caption{Information \req{statinfo} (squares) and percentage $r_{\pi}$ of games played by the
fittest agent (circles) over time in model \req{bechmod}.}\label{ingrofig}
\end{figure}
\begin{figure}
\centerline{\epsfxsize=3.0in \epsfbox{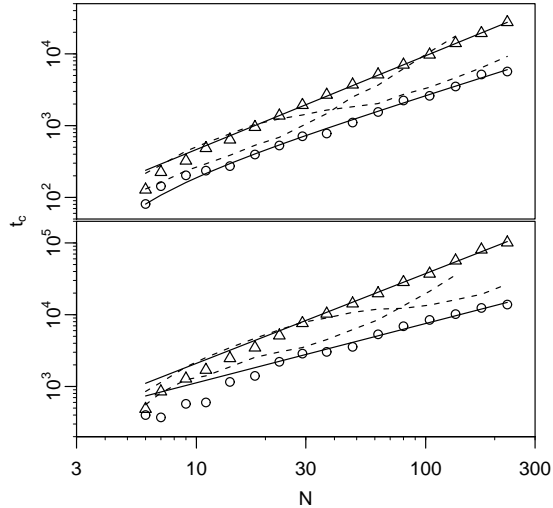}} 
\vspace{-1cm}
\caption{Average number of games needed to find the best object with probability
$0.6$ in model \req{bechmod} with $\pi =0.60$ in the upper and $\pi
=0.55$ in the lower panel respectively. Symbols report simulation results for 
a repeated Round Robin tournament (triangles) and for the min-ent
method (circles). The dashed lines are for the {\it ran--fil} with
$\eta=1,7$, where $\eta=7$ gives the best performance for large $N$.
In the upper panel the min-ent method is fitted with a linear fit,
the robin tournament with a power law fit with exponent $1.5$. In the
lower panel the fitting exponents are $0.82$ respectively $1.25$.} \label{fig_me_1}
\end{figure}
We tested the min-ent method on the
model \req{bechmod}. First we verified that it converges,
i.e. entropy \req{ch1} goes to zero. In figure \ref{ingrofig}
we show that it is actually the case:
the information gain approaches $1$,
\be\label{statinfo}
I_{W(t)}=1-H\left(\psi|W(t)\right)/H_{max} \to 1,
\ee
as we employ more resources ($t \to \infty$),
and so does the percentage $r_{\pi}$ of games played by the
fittest agent.
Then we compared the performance of the min-ent method 
with that of the ran-fil method and with 
repeated RRs; results are shown in figure \ref{fig_me_1}.
The total number of comparisons $t_c$ needed to pick
the best item with a given probability ($0.6$ in the figure)
seems to scale linearly with the number of items $N$ using
the min-ent method, while it is a super-linear power law for RRs.
The ran-fil method, although less performing than the min-ent,
clearly outperforms Round Robins. This last result is particularly
promising, since it is widely applicable to real filters.

\section*{Conclusions}
We have analyzed different methods of selecting a set of $N$ objects
by means of $n_c$ pairwise comparisons. In particular we focused on
the amount of resources one has to spend in order to select the fittest.
We stated that the best one can
obtain with repeated round robins with successive elimination, for a wide variety
of probabilistic models, is $n_c\propto N^2 \log N$. Then we introduced
two new methods of performing the selection. 
Both ones give better results than repeated RRs in a worst-case
test model: with the min-ent, based on information maximization, we
obtained $n_c \propto N$. The ran-fil method has been shown to
outperform round robins without requiring any previous knowledge
of the underlying model. Its basic principles are to 
keep an almost constant selecting power at each round and to gradually
eliminate losers. We believe they
constitute a useful proposal for improving tournament design, with
particular reference to ranking methods of Internet Search Engines.

\section*{Acknowledgments}
We would like to acknowledge Joseph Wakeling, Andrea Capocci
and Yi-Kuo Yu for useful discussions. This work has been
supported by the Swiss National Fund under grant number 2051-67733.


%
\end{document}